\documentclass[12pt]{iopart}
\usepackage{iopams}
\usepackage{graphicx}
\usepackage{amssymb}
\usepackage{amstext}
\usepackage{latexsym}
\usepackage{ulem}
\usepackage{color}
\usepackage{subfigure}

\begin{document}

\title{Spin-wave excitations in presence of nanoclusters of magnetic impurities}

\author{Akash Chakraborty$^{1,2,3}$, Paul Wenk$^{3}$, Stefan Kettemann$^{2,4}$, Richard Bouzerar$^{5}$, Georges Bouzerar$^{5}$}

\address{
$^1$ Institut f\"ur Theoretische Festk\"orperphysik, Karlsruhe Institute of Technology, 76128 Karlsruhe, Germany
}
\ead{akash.chakraborty@physik.uni-regensburg.de}

\address{
$^2$ School of Engineering and Science, Jacobs University Bremen, Campus Ring 1, 28759 Bremen, Germany
}

\address{
$^3$ Institut I - Theoretische Physik, Universit\"at Regensburg, 93040 Regensburg, Germany
}

\address{
$^4$ Division of Advanced Materials Science, Pohang University of Science and Technology (POSTECH), Pohang 790-784, South Korea
}

\address{
$^5$ Institut N\'eel, CNRS, D\'epartement MCBT, 25 avenue des Martyrs, B.P. 166, 38042 Grenoble Cedex 09, France
}


\begin{abstract}
Nanoscale inhomogeneities and impurity clustering are often found to drastically affect the magnetic and transport properties in disordered/diluted systems, giving rise to rich 
and complex phenomena. However, the physics of these systems still remains to be explored in more details as can be seen from the scarce literature available. We 
present a detailed theoretical analysis of the effects of nanoscale inhomogeneities on the spin excitation spectrum in diluted magnetic systems. The calculations are 
performed on relatively large systems (up to $N$=$66^3$). It is found that even low concentrations of inhomogeneities have drastic effects on both the magnon density of states 
and magnon excitations. These effects become even more pronounced in the case of short ranged magnetic interactions between the impurities. In contrast to the increase of 
critical temperatures $T_C$, reported  in previous studies, the spin-stiffness $D$ is systematically suppressed in the presence of nanoscale inhomogeneities. Moreover $D$ is 
found to strongly depend on the inhomogeneities' concentration, the cluster size, as well as the range of the magnetic interactions. 
The findings are discussed in the prospect of potential spintronics applications.
We believe that this detailed numerical work could initiate future experimental studies to probe this rich physics with the most appropriate tool, 
Inelastic Neutron Scattering (INS). 
\end{abstract}


\maketitle

\section{Introduction}

\ Disordered magnetic systems, such as transition metal alloys\cite{moruzzi,abrikosov,james}, diluted magnetic semiconductors (DMSs)\cite{junqwirth,satormp,timm}, 
diluted magnetic oxides (DMOs)\cite{fukumara,chambers}, and manganites\cite{salamon,goto,dagotto} have continued to attract considerable 
attention ever since their respective discoveries. The prospects of these materials for novel spintronic devices led to a plethora of work from both the experimental as well as 
the theoretical point of view. For a long time the major focus of spintronics research was mostly on homogeneously diluted systems (with magnetic impurities distributed 
randomly on the host lattice), which were believed to give rise to the much coveted room-temperature ferromagnetism. For this purpose 
huge efforts to grow systems as ``clean'' as possible were made. 
However, in contrast to the common belief, this does not systematically appear to be the optimal route to room-temperature ferromagnetism. For instance, in homogeneously or 
well-annealed Mn doped III-V DMSs it has been impossible, so far, to go beyond the critical temperature of 140 K for 5\% Mn\cite{edmonds,ku,richard1}.

\ These findings led researchers to look into other possible avenues in the ultimate quest for the room-temperature phenomenon. Transmission electron 
microscopy (TEM) experiments revealed the presence of coherent Mn-rich spherical nanocrystals in (Ga,Mn)As\cite{moreno,yokoyama}, which exhibited a Curie temperature of 
$\sim$360 K. 
A careful micro-structure analysis revealed the cluster structure to be coherent with that of the host matrix (zinc-blende type), and this ruled out the possibility 
of any inter-metallic precipitates or phase separation in these materials. 
Similar nanoscale clusters were also reportedly observed in other diluted materials like (Ge,Mn)\cite{xiu,bougeard,park} and (Zn,Co)O\cite{jedrecy}, exhibiting critical 
temperatures in excess of 300 K. A recent TEM-based study on (In,Fe)As\cite{tanaka} has also reported the presence of magnetic clusters, which were neither 
metallic Fe nor inter-metallic FeAs compounds.
The nanoscale spinodal decomposition into regions with high and low concentration of magnetic ions was speculated to be the possible reason for the 
apparently high Curie temperatures in these compounds. Nanoscale inhomogeneities have also often been detected in other families of compounds.  In manganites, inhomogeneities 
arise due to the interplay between the charge, spin, orbital, and lattice degrees of freedom, leading to the coexistence of metallic and insulating phases. For example, 
scanning tunneling spectroscopy (STS) images of La$_{1-x}$Ca$_x$MnO$_3$ revealed a clear phase separation just below the critical temperature \cite{fath}. It is widely 
believed that the phase separation is at the origin of the well known colossal magneto-resistance (CMR) effect\cite{helmholt1993,jin1994}. In  high $T_C$ superconductors 
experimental studies often indicate  the crucial role of inhomogeneities\cite{pan2001,wise2009,parker2010} . For example in \cite{parker2010}, it was found that 
regions with weak superconductivity can persist to higher temperatures if bordered by regions of strong superconductivity. The possibility of 
increasing the maximum transition temperature by controlled distribution of the dopants has also been suggested. 

\ Despite the existence of numerous experimental studies, the effects of nanoscale inhomogeneities in disordered and diluted systems still remain largely unexplored 
on the theoretical front. Because of the large supercells required density functional theory (DFT) based calculations for inhomogeneous systems are considerably difficult.
For the same reason the essentially exact Monte Carlo studies are unfortunately incapable to deal with these 
systems. Moreover the crucial importance of both thermal and transverse fluctuations calls for the inevitable need of an exact treatment of the disorder effects, which implies 
a real space treatment. Recently using a real-space RPA approach, able to handle considerably large system sizes, it has been demonstrated that nanoscale 
inhomogeneities can lead to a dramatic increase of the Curie temperatures\cite{akash2012_1}. The question which naturally arises is how impurity clustering 
affects the spin excitation spectrum? Indeed the spin-wave excitations provide valuable insight into the underlying spin dynamics of the system.  Inelastic neutron scattering 
is a powerful experimental tool in this context, as the dispersion relation and the magnon damping can be measured directly and accurately. In contrast to non-dilute systems, 
such as manganites\cite{motome2005,zhang2007}, cobaltites\cite{louca2003,hoch2004}, multiferroics\cite{poienar2010,haraldsen2010}, pnictides\cite{maier2008,ewings2008}, etc., 
one can find very few detailed studies\cite{uemura1986} devoted to spin dynamics in dilute magnetic systems. Note that till now no experimental data has been 
reported for spin excitations in III-V DMSs, such as (Ga,Mn)As.

\begin{figure}[htbp]\centerline
{\includegraphics[width=5.0in,angle=0]{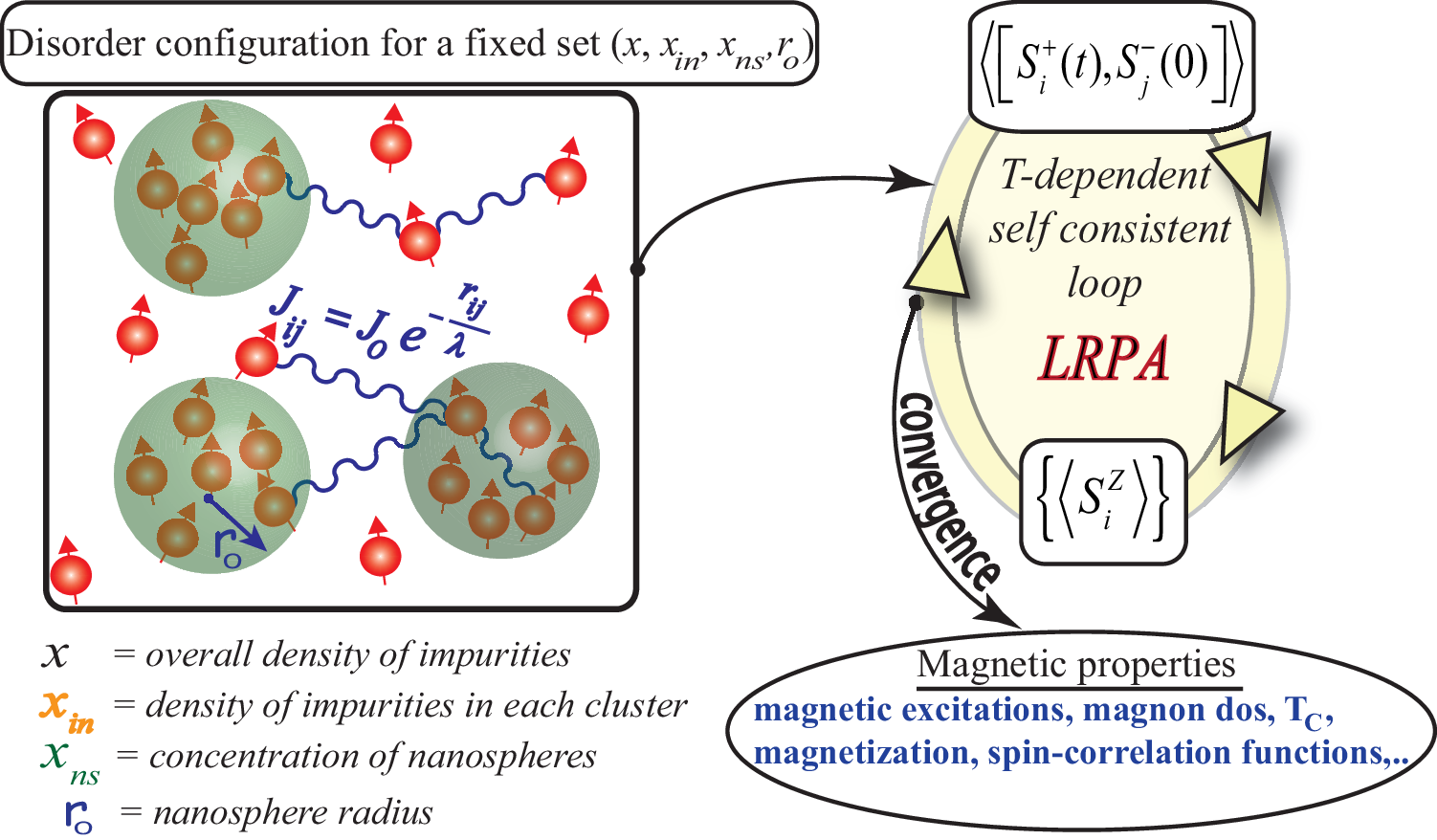}}
\caption{Schematic representation of a disorder configuration, illustrating the model parameters, and the self-consistent local RPA method.
}
\label{fig1}
\end{figure}

\section{Theoretical approach and the model}

\ The aim of the current work is to provide such a detailed theoretical account of the spin excitation spectrum of diluted magnetic systems in the presence of nanoscale 
inhomogeneities. We show that nanoclusters of magnetic impurities can have drastic effects on the magnon density of states (DOS), 
the dynamical spectral function, as well as the spin-stiffness in these systems, when compared to the homogeneously diluted case. 
We recall that the homogeneous case refers to a random and uncorrelated distribution of the impurities without any nanoclusters. In our calculations, 
we have assumed for the lattice
a simple cubic structure with periodic boundary conditions. In order to avoid additional parameters, the total concentration of impurities 
in the whole system is fixed to $x$=0.07. The inhomogeneities are assumed to be spherical of radii $r_0$ and the concentration inside each of these nanospheres 
is denoted by $x_{in}$. For simplicity we also fix $x_{in}$=0.8 for all cases considered in this study. 
This implies that for a size $r_0$=$2a$ (where $a$ is the lattice spacing) each nanosphere contains 26 impurities for this $x_{in}$. 
We define the concentration of nanospheres in the system as  
$x_{ns}=N_{S}$/$N$, where $N_{S}$ is the total number of sites included in all the nanospheres and $N$=$L^3$ is the total number of sites in the lattice. A variable 
$P_{N}$=$(x_{in}/x)x_{ns}$, is used to represent the total fraction of impurities contained within the nanospheres. 
For a particular disorder configuration the nanospheres are chosen in a random manner on the lattice, the only restriction imposed is to 
avoid any overlapping between them. 
The overlapping is avoided to ensure that we always have well defined clusters, as the aim is to clearly identify the effects of a given kind of inhomogeneities 
on the magnetic properties. Nevertheless, even without this restriction we have checked that the effects are negligible on the results, and the conclusions in general are not 
affected. A schematic illustration of a typical disorder configuration and the methodology is given in Figure~\ref{fig1}. This model can be useful for 
a qualitative study of materials such as (Ga,Mn)As or (Zn,Co)O, where coherent spherical nanoclusters with diameters in the range of 5-50 nm were 
reported\cite{moreno,yokoyama,jedrecy}.

We start with a Hamiltonian describing $N_{imp}$ interacting spins randomly distributed on a lattice of $N$ sites, which is given by the diluted random Heisenberg model
\begin{eqnarray}
H_{Heis}=-\sum_{i,j} J_{ij}p_{i}p_{j} {\bf S}_{i}\cdot{\bf S}_{j}
\label{Hamiltonian}
\end{eqnarray}
where the sum $ij$ runs over all sites and the random variable $p_i$ is 1 if the site is occupied by an impurity, otherwise it is 0. The above Hamiltonian (1) is treated 
within the self-consistent local random phase approximation (SC-LRPA), which is a semi-analytical approach based on finite temperature Green's functions. Within this approach, 
the retarded Green's functions are defined as $G^c_{ij}(\omega)=\int_{-\infty}^{+\infty}G^c_{ij}(t)e^{i\omega t}dt$, where 
$G^c_{ij}(t)=-i\theta(t)\langle[{\bf S}_i^+(t),{\bf S}_j^-(0)]\rangle$, ($\langle ...\rangle$ denotes the thermal average). This describes the transverse spin fluctuations and the 
index ``$c$'' denotes the disorder configuration. Note that the notation $G^c_{ij}(\omega)$ is equivalent to $G^c(r_i,r_j,\omega)$. 
After performing the Tyablicov decoupling of the higher-order Green's functions which appear in the equation of motion of 
$G^c_{ij}(\omega)$ one gets,
\begin{eqnarray}
G_{ij}^c(\omega)=\sum_\alpha \frac{2\langle{S_j^z}\rangle}{\omega-\omega_\alpha^c+i\epsilon}
\langle i|\Psi_\alpha^{R,c}\rangle \langle\Psi_\alpha^{L,c}|j\rangle
\label{GF}
\end{eqnarray}
where $\langle{S_j^z}\rangle$ are the local magnetizations which have to be calculated self-consistently. $|\Psi_\alpha^{R,c}\rangle$ and $|\Psi_\alpha^{L,c}\rangle$ are 
respectively the right and left eigenvectors, associated with the same eigenvalue $\omega^c_{\alpha}$ of the effective Hamiltonian ${\bf H}^c_\text{eff}$, whose matrix 
elements are given by
\begin{eqnarray}
({{\bf H}_\text{eff}^c})_{ij}=-\langle{S_i^z}\rangle J_{ij}+ \delta_{ij}\sum_{l}\langle{S_l^z}\rangle J_{lj}
\label{heff}
\end{eqnarray}
where $c$ denotes the configuration, $i,j,l$ are the site indices, and $z$ is the polarisation direction. 
At finite temperature, ${\bf H}^c_\text{eff}$ is non-Hermitian (real and not necessarily symmetric) but it is bi-orthogonal\cite{varma}. 
The property of bi-orthogonality requires the calculation of the left and right eigenvectors, both associated with the same eigenvalue. 
However, since in the present case all calculations are performed at $T$=0 K, the matrix ${\bf H}^c_\text{eff}$ is real symmetric, and the left and right eigenvectors are identical. 
Also, since at $T$=0 K the ground state is fully polarised, all $\langle{S_i^z}\rangle$'s =$S$ in this case. 
For more details on the SC-LRPA approach one can see for example \cite{satormp,georges2005epl,georges2007}. The SC-LRPA was previously successfully implemented to 
calculate the magnetic excitation spectra in the nearest-neighbor diluted Heisenberg model\cite{akash2010} as well as in the case of optimally annealed 
(Ga,Mn)As\cite{akash2011}. In the latter case, an especially good agreement with experiments\cite{goennenwein2003,sperl2008} was also found. 

\ Now it is well known from first principles based studies as well as model calculations that the exchange couplings in III-V DMS compounds are relatively short-ranged, 
non-oscillating, and almost exponentially decaying in nature\cite{akash2011,bergqvist,satoprb}. 
This was the primary motivation to assume generalized couplings of the form 
$J_{ij}$=$J_{0}$exp(-$\mid$$ \textbf r_i$-$\textbf r_j $$\mid$/$\lambda)$, where $\lambda$ is the damping parameter. For this damping parameter, we focus here on two 
particular values, $\lambda=a$ and $a/2$ ($a$ is the lattice spacing), corresponding to relatively long-ranged and short-ranged couplings respectively. 
It is worth noting that for about 5\% Mn-doped GaAs, a fit of the \textit{ab initio} exchange couplings leads to a value of $\lambda$ of the order of $a/2$. 
Only recently it was shown that just varying $\lambda$ within this scale can give rise to rather spectacular effects on the Curie temperatures in these inhomogeneous diluted 
systems\cite{akash2012_1}. 
Note that the nature of the magnetic couplings could also depend on the concentration of the clusters and/or the density of impurities inside. However, computing 
these couplings involves extensive calculations, systematic sampling over disorder, and especially diagonalizing very large matrices, which is beyond the scope of the present 
work. Our current goal is to provide a detailed analysis of the effects of nanoscale inhomogeneities, in the presence of these generalized interactions, 
on the dynamical properties of diluted systems, but the conclusions drawn are general.

\section{Numerical results and discussion}

\begin{figure}[htbp]\centerline
{\includegraphics[width=4.3in,angle=0]{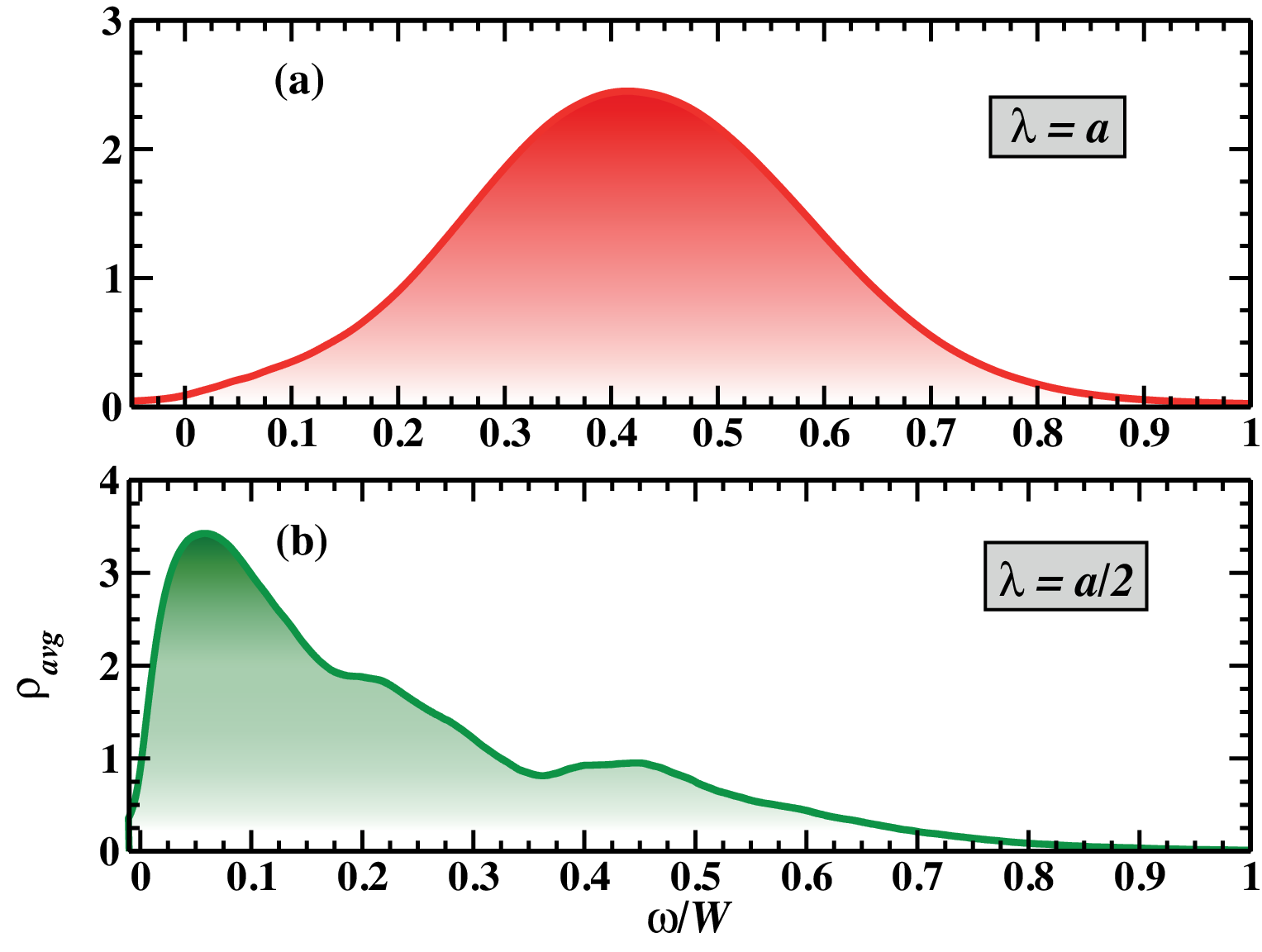}}
\caption{Average magnon DOS corresponding to the homogeneous case for (a) $\lambda$=$a$, and (b) $\lambda$=$a/2$. 
The average concentration is $x$=0.07. The $x$ axis is in units of $\omega/W$, where $W$ is the magnon spectrum bandwidth. $W$$\approx$4$J_{0}$ and 0.8$J_{0}$, for 
$\lambda$=$a$ and $a/2$ respectively. The system size is $N$=56$^3$.
}
\label{fig2}
\end{figure} 

\ We first start with the calculation of the magnon DOS for the homogeneously diluted case. The average magnon DOS is given by 
$\rho_{avg}(\omega)=\frac{1}{N_{imp}}\sum_{i}\langle\rho_{i}^{c}(\omega)\rangle_c$, 
where $\rho_{i}^{c}(\omega)$ is the local magnon DOS, which reads $\rho_{i}^{c}(\omega)=-\frac{1}{2\pi\langle{S_i^z}\rangle} {\rm Im} G_{ii}^{c}(\omega)$. 
In what follows $\langle...\rangle_c$ denotes the average over disorder configurations. 
Figure~\ref{fig2} shows $\rho_{avg}$ as a function of the energy $\omega$, 
for the two different values of $\lambda$ mentioned above. The average over disorder is performed for a few hundred configurations (typically 200) 
in all the following calculations, although it was found that about 100 configurations are sufficient for the sampling and the results converge beyond 
that. $\rho_{avg}$ is found to exhibit a regular Gaussian-like shape for the case of $\lambda$=$a$. The broad peak is located at 0.42$W$ with a half-width 
of about 0.36$W$ ($W$ is the magnon excitation bandwidth). For longer ranged couplings, $\rho_{avg}$ remains essentially similar to that of  $\lambda$=$a$. On the other hand, 
for short-ranged couplings ($\lambda$=$a/2$), $\rho_{avg}$ has a more irregular and richer structure. The peak in $\rho_{avg}$ is now located at much lower energy, 
0.06$W$, and a clear long tail extending toward the high energies with multiple shoulders appears. 
These irregular features can be attributed to regions of impurities which are weakly connected to each other due to the short-range nature of the 
interactions. Effectively they behave as isolated regions which have their own eigenmodes, and these in turn contribute to the irregular shoulders 
appearing in the DOS. These shoulders become even more pronounced for shorter ranged interactions. In the case of $\lambda$=$a$ these shoulders are absent 
because of the longer range nature of the couplings.  
It is interesting to note that a similar kind of magnon DOS (as in Figure~\ref{fig2}(b)) was obtained in the case of (Ga,Mn)As\cite{akash2011}. However, 
in \cite{akash2011} the exchange couplings used had been directly calculated from the ``$V$-$J$'' model. It has been found that $\lambda$=$a/3$ provides a very good 
fit for these couplings.

\begin{figure}[t]\centerline
{\includegraphics[width=4.5in,angle=0]{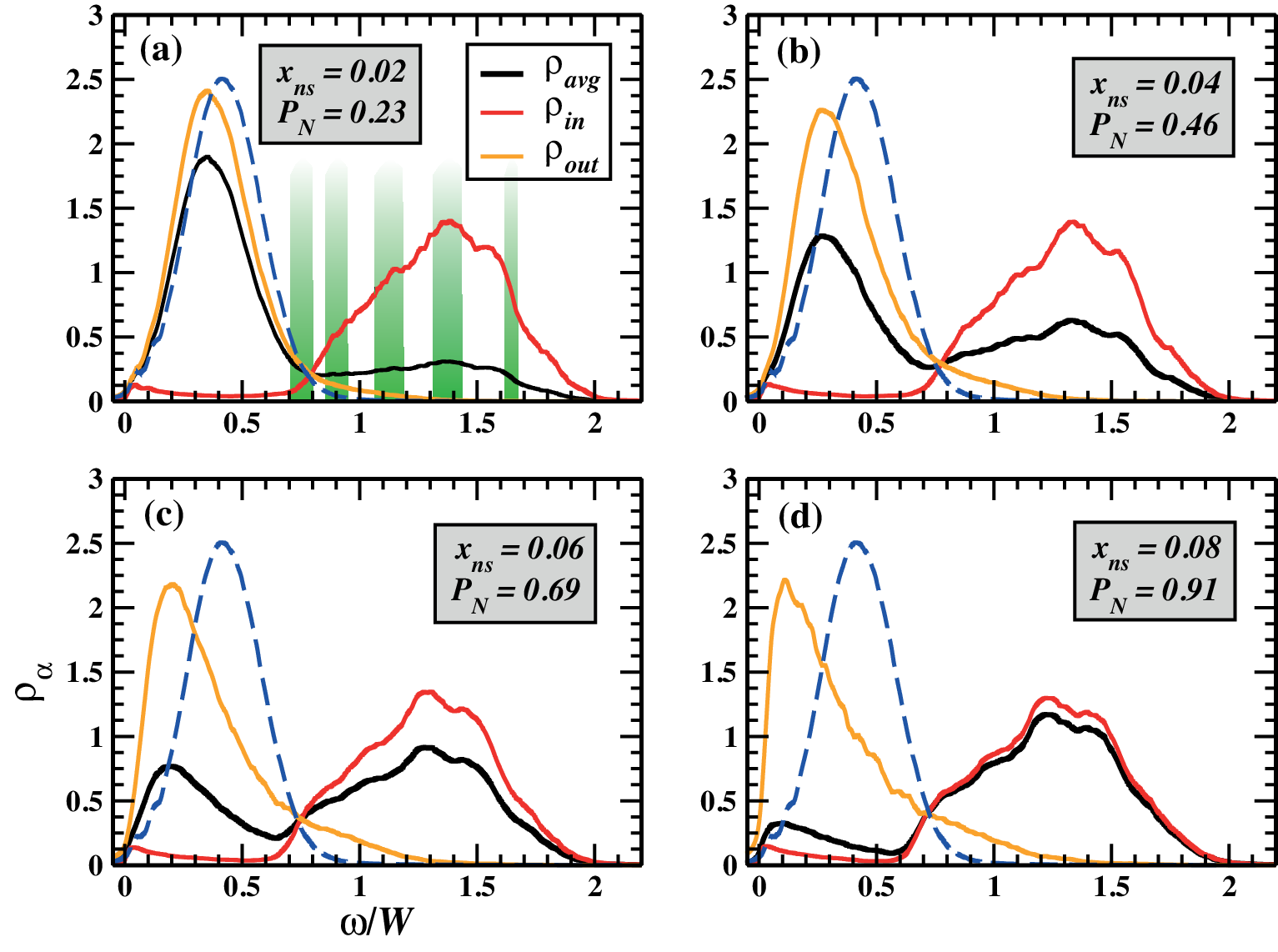}}
\caption{ $\rho_{\alpha}$ denotes the average magnon DOS ($\rho_{avg}$), local magnon DOS inside ($\rho_{in}$), and outside the nanospheres ($\rho_{out}$), 
for four different $x_{ns}$. $P_N$ is the percentage of total impurities inside the nanospheres. 
The blue dashed curves represent the homogeneous $\rho_{avg}$ from Figure 2(a). The shaded regions in (a) correspond to the calculated eigenmodes of a single isolated nanosphere. 
The parameters are $\lambda$=$a$, $r_0$=$2a$, and $x_{in}$=0.8. The $x$ axis is in units of $\omega/W$, where $W$$\approx$4$J_{0}$ is homogeneous magnon bandwidth. 
} 
\label{fig3}
\end{figure} 

\ In order to analyze the effects of inhomogeneities, we calculate in addition to $\rho_{avg}$, the local magnon DOS inside ($\rho_{in}$) and outside ($\rho_{out}$) 
the nanospheres. Unlike $\rho_{avg}$, the local DOS contains information on the nature of the magnon states, whether they are extended or localized. One can 
also calculate the typical DOS (which is essentially the geometric mean of the local DOS) which provide a direct access to the ``mobility edge'' separating the localized modes 
from the extended ones. This can be the subject of a future detailed study. For the local DOS, 
we consider the particular case of nanospheres with fixed radii $r_0$=$2a$, concentration inside $x_{in}$=0.8, and four different concentrations of nanospheres $x_{ns}$. 
The results for $\lambda$=$a$ are depicted in  Figure~\ref{fig3}. Let us first focus on $\rho_{avg}$. 
From Figure~\ref{fig3}(a), we immediately notice that a relatively small concentration of inhomogeneities ($x_{ns}$$\sim$0.02) causes a significant change in the magnon DOS. 
Indeed, in comparison to the homogeneous case, the excitations spectrum bandwidth is now doubled, and $\rho_{avg}$ has a bimodal structure, with a broader peak at higher 
energies. With increasing $x_{ns}$, we observe a gradual transfer of weight from the low to high energy peak. The low energy peak shifts to 
smaller energies which is consistent with the decrease in the concentration of impurities outside the nanospheres. 
In order to have a better understanding of the features seen in $\rho_{avg}$, we now analyze $\rho_{in}$ and $\rho_{out}$.
We observe that $\rho_{in}$ remains unchanged in all cases and exhibit a very small weight from 0 to 0.7$W$. Thus the high energy peak seen in $\rho_{avg}$ can clearly be 
attributed to the nanocluster modes. A careful analysis of a single isolated cluster reveals that the first non-zero eigenmodes are located at 0.7$W$, which explains the very 
small weight in $\rho_{in}$ below this value. Note that in Figure 3(a), the shaded regions correspond to the discrete spectrum of an isolated single nanosphere, which is calculated 
over a few hundred configurations (random position of the impurities inside the nanosphere). 
The weak variation of $\rho_{in}$ with respect to $x_{ns}$, indicates that the disappearance of the discreteness in $\rho_{in}$ (as seen in the isolated nanosphere spectrum)
results mainly from the interactions between the cluster impurities and those outside.  The above discussion of $\rho_{avg}$ and $\rho_{in}$ explains naturally the behavior 
of $\rho_{out}$. In the case of more extended couplings, it is expected that  (i) $\rho_{avg}$ loses progressively the bimodal nature, (ii) the pseudo-gap in $\rho_{in}$ at 
low energies is filled gradually, and (iii) the second peak in $\rho_{in}$ becomes narrower and shifts to higher energies with respect to the spectrum of a single isolated cluster.

\begin{figure}[t]\centerline
{\includegraphics[width=4.5in,angle=0]{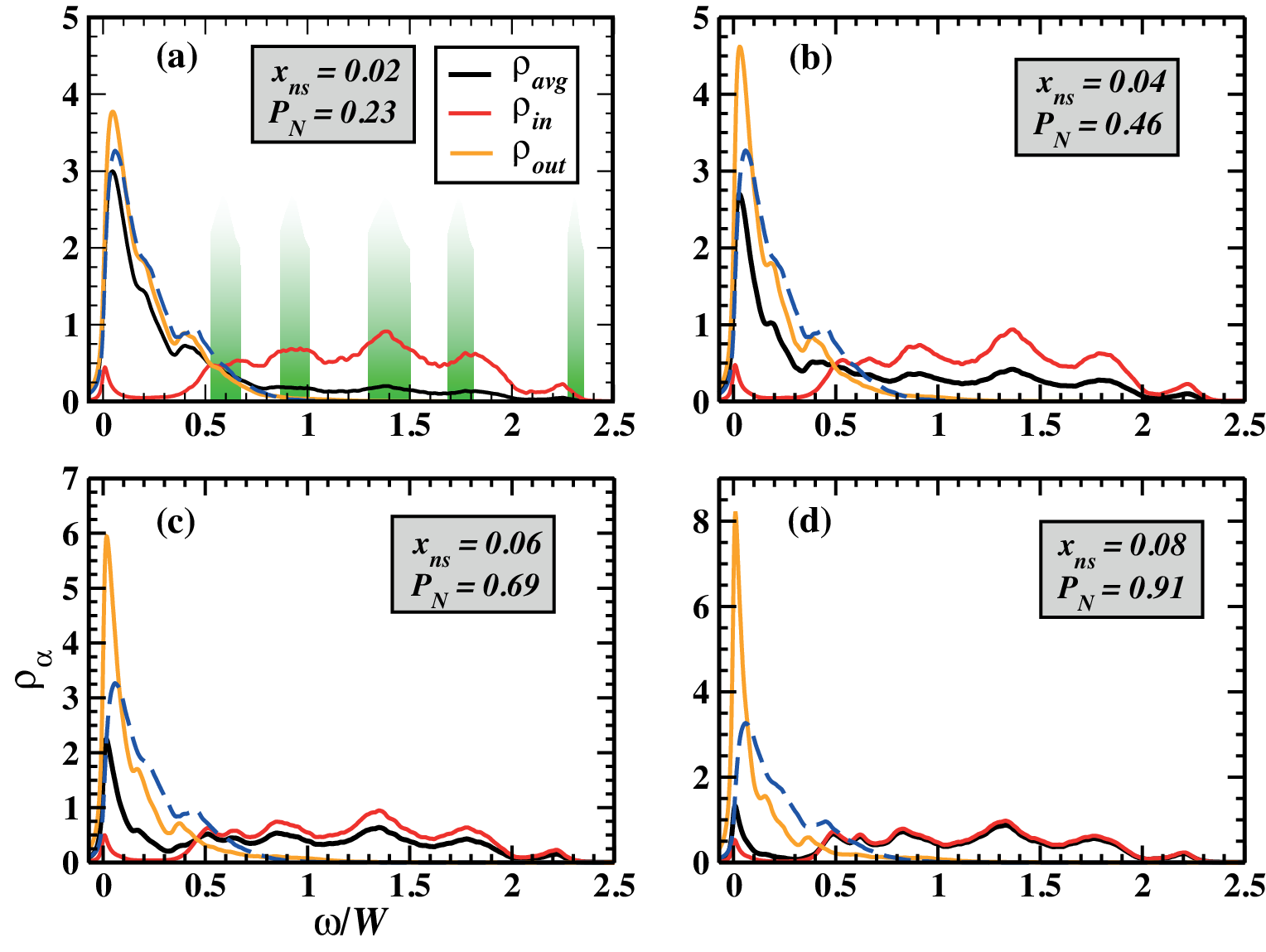}}
\caption{ $\rho_{avg}$, $\rho_{in}$, and $\rho_{out}$ for $\lambda$=$a/2$, for four different $x_{ns}$. The blue dashed curves represent the homogeneous 
$\rho_{avg}$ from Figure 2(b). The shaded regions in (a) correspond to the calculated eigenmodes of a single isolated nanosphere. The other parameters being same as 
in Figure~\ref{fig3}. The homogeneous magnon bandwidth is $W$$\approx$0.8$J_{0}$. 
}
\label{fig4}
\end{figure}

\ We now discuss the effects of short ranged interactions ($\lambda$=$a/2$) on the average and local magnon DOS. The results are shown in Figure~\ref{fig4}.
The magnon bandwidth increases by 250\% with respect to that of the homogeneous system. As seen before, we observe a clear transfer of weight in $\rho_{avg}$, from the 
low to the higher energies with increasing  $x_{ns}$. In contrast to the bimodal nature observed for $\lambda$=$a$, $\rho_{avg}$ now exhibits a long wavy tail, 
extending toward higher energies. 
This wavy tail is actually attributed to the cluster impurities and is related to the nature of $\rho_{in}$, which is discussed in the following. 
$\rho_{in}$ shows (i) a clear multiple peak structure now, (ii) is independent of $x_{ns}$, and (iii) a well defined gap of 
approximately 0.5$W$ is observed. The reasons for the appearance of these multiple peaks in $\rho_{in}$ are the enhanced discreteness (larger sub-gaps) 
of the eigenmodes of the single isolated nanosphere and the reduced interactions of the cluster impurities with those outside. 
Concerning $\rho_{out}$, besides a shift to lower energies as seen for $\lambda$=$a$, we now observe that the peak becomes narrower with increase in $x_{ns}$.
(The latter feature was absent for the long ranged couplings). The reason for this is with increasing $x_{ns}$, the concentration of impurities outside decreases and 
the effective interactions between them become weaker. This effect will be even more pronounced for shorter ranged couplings. 
Even though the couplings are comparable, drastic changes between Figure~\ref{fig3} and Figure~\ref{fig4}, shows that $\lambda$=$a$ corresponds 
to the intermediate range couplings and $\lambda$=$a/2$ definitely to the short range regime.

\begin{figure}[htbp]
\centering
\subfigure{
\includegraphics[width=2.4in,angle=-90]{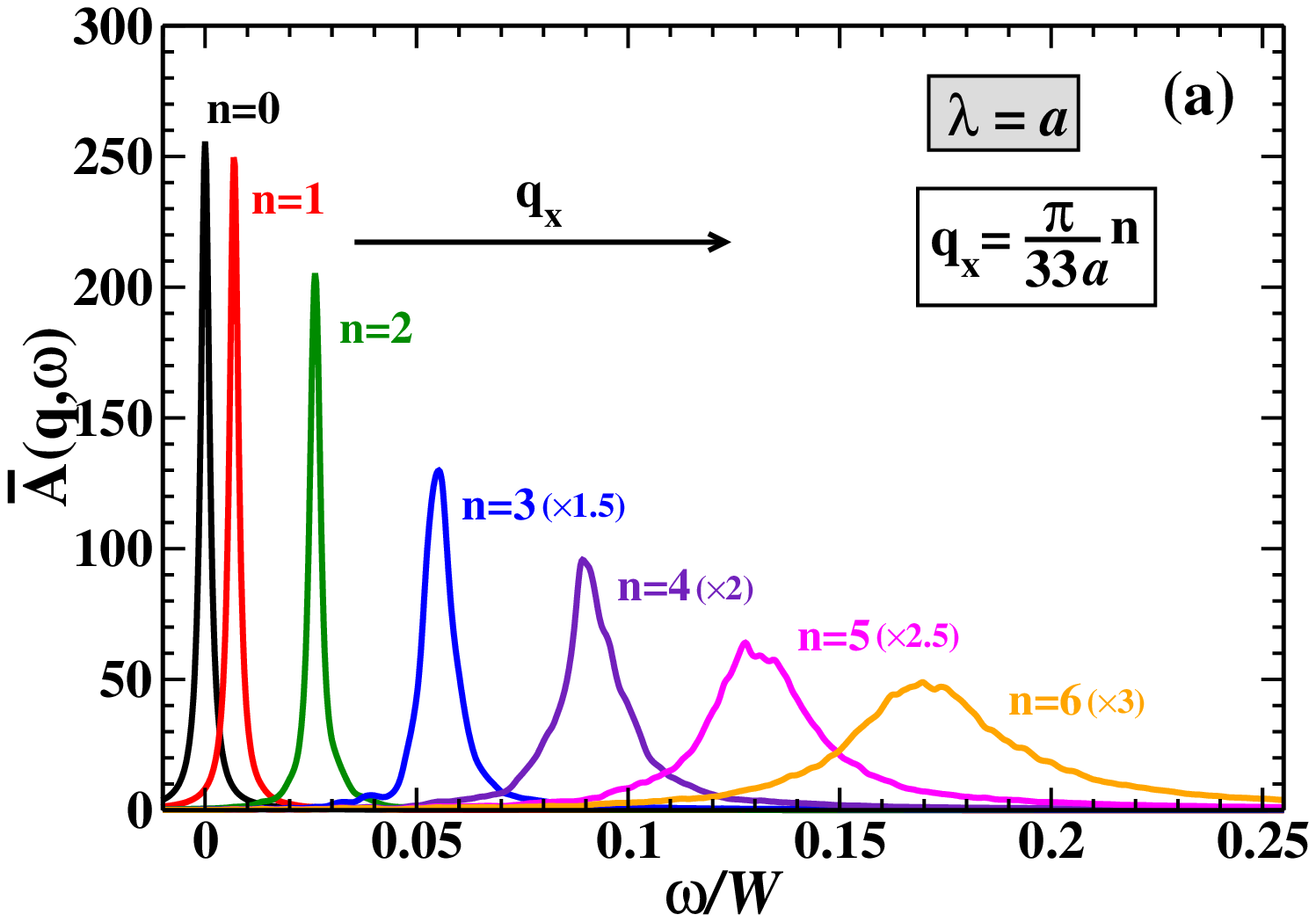}
\label{fig5a}
}
\hspace{-1.22cm}
\subfigure{
\includegraphics[width=2.4in,angle=-90]{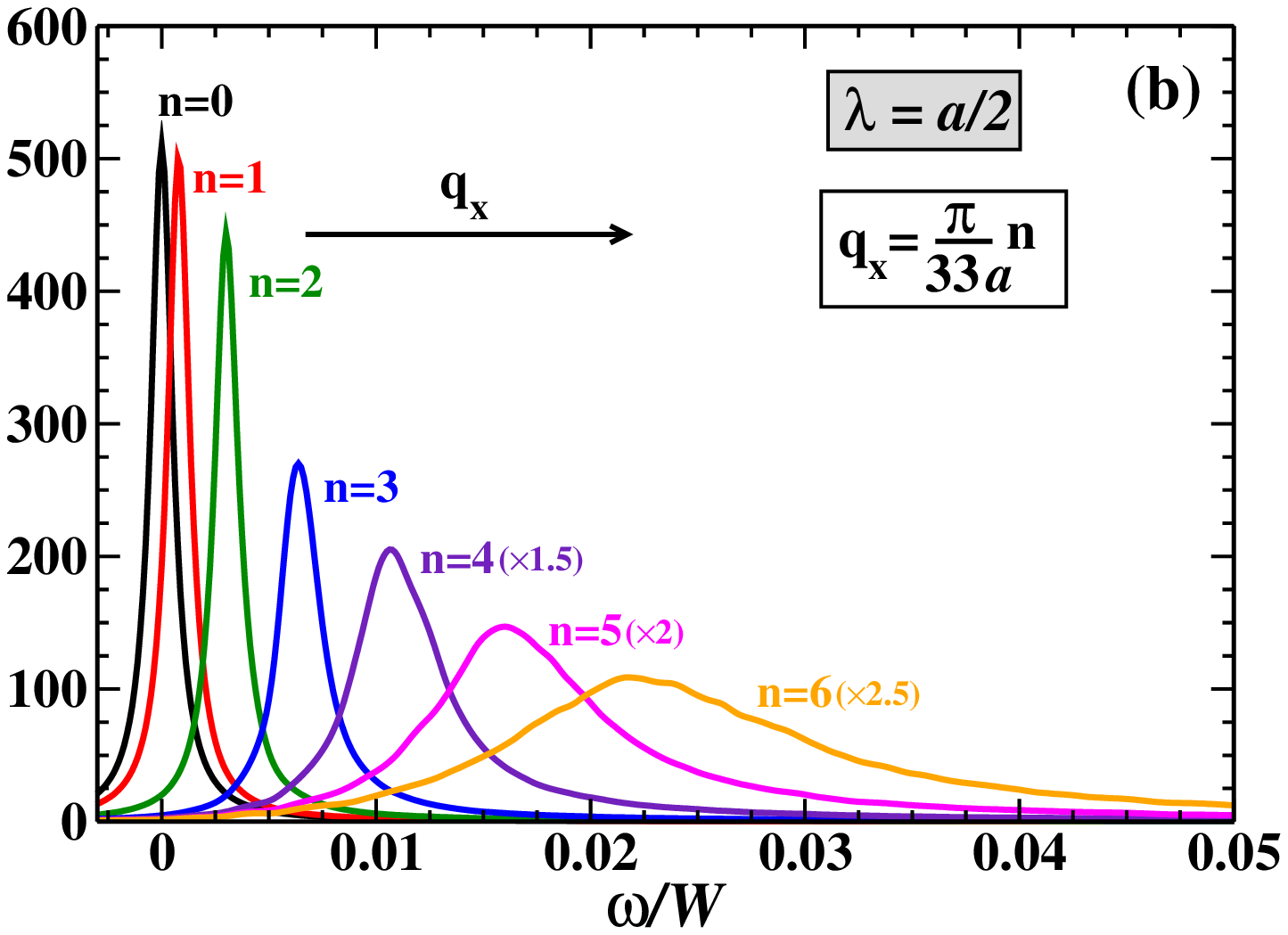}
\label{fig5b}
}
\label{fig5}
\caption[Optional caption for list of figures]{
 Average spectral function $\bar{A}({\bf q},\omega)$ as a function of the energy in the [1 0 0] direction for different values of $q_x$, corresponding 
to the homogeneous case for (a) $\lambda$=$a$, and (b) $\lambda$=$a/2$. The energy axis ($x$-axis) is in units of $\omega/W$. The system size is $N$=66$^3$. (The 
intensity of the peaks have been multiplied by the factors indicated in the parentheses).
}
\end{figure}

\ We propose now to focus on the low energy excitation spectrum in these systems. For this purpose we evaluate the dynamical spectral function, which provides deeper insight 
into the underlying spin dynamics. This physical quantity can be directly and accurately probed by inelastic neutron scattering (INS) experiments. 
It is defined by
$\bar{A}({\bf q},\omega) \equiv \langle A^c({\bf q},\omega)\rangle_c=-\left\langle \frac{1}{2\pi\langle\langle{S^z}\rangle\rangle} {\rm Im} G^c({\bf q},\omega)\right\rangle_c$, 
where $\langle\langle{S^z}\rangle\rangle=\frac{1}{N_{imp}}\sum_{i}\langle{S_i^z}\rangle$ is the total average magnetization over all spin sites, and 
$G^c({\bf q},\omega)$ is the Fourier transform of $G_{ij}^c(\omega)$ given in equation (2). The averaged dynamical spectral function is evaluated from the following 
\begin{eqnarray}
\bar{A}({\bf q},\omega) = \left\langle \sum_\alpha A_{\alpha}^c({\bf q})\delta(\omega-\omega^c_{\alpha})\right\rangle_c
\label{Aqw}
\end{eqnarray}
where 
\begin{eqnarray}
 A_{\alpha}^c({\bf q})= \frac{1}{N_{imp}} \sum_{ij} \lambda_j \langle i|\Psi_\alpha^{R,c}\rangle \langle\Psi_\alpha^{L,c}|j\rangle e^{i{\bf q(r}_i-{\bf r}_j)}
\label{Acq}
\end{eqnarray}
where $\lambda_j=\frac{\langle{S_j^z}\rangle}{\langle\langle{S^z}\rangle\rangle}$ is the temperature dependent local parameter. Note that at $T=0$ K, all $\lambda_j$'s = 1.

\ First we discuss the average spectral function for the homogeneously diluted systems. Figures~\ref{fig5a} and \ref{fig5b} show  $\bar{A}({\bf q},\omega)$, for $\lambda$=$a$ 
and $a/2$ respectively, as a function of the energy for different values of the momentum {\bf q} in the [1 0 0] direction. First in both cases, as {\bf q} increases the peaks 
become broader and more asymmetric with a tail extending toward higher energies. Well-defined excitations exist only for relatively small values of the momentum, beyond 
$q_{x}a \approx 0.24\pi$ (n=8) no well-defined magnons exist. 
The increase in asymmetry with increasing {\bf q} (i.e. shorter wavelength) is due to the decay of the collective spin waves into localized excitations. A 
considerable softening of the spin waves will be observed on approaching the Brillouin zone boundary, which is due to the localized nature of the high-energy magnon modes in 
this region. 
However, for $\lambda$=$a$, the  well-defined excitations persist up to energy values of about 0.25$W$, whilst for $\lambda$=$a/2$ the excitations reach only up to 
0.035$W$, where $W$ is the $\lambda$-dependent magnon spectrum bandwidth. 
This difference in the range of energies, by one order of magnitude, can be attributed to the decrease in the range of the effective interactions. This also 
demonstrates, once again, the profound effects observed on going from $\lambda$=$a$ to $a/2$. We remind that $W$$\approx$ 4$J_0$ and 0.8$J_0$, for $\lambda$=$a$ and 
$a/2$, respectively. 

\begin{figure}[t]\centerline
{\includegraphics[width=4.2in,angle=-90]{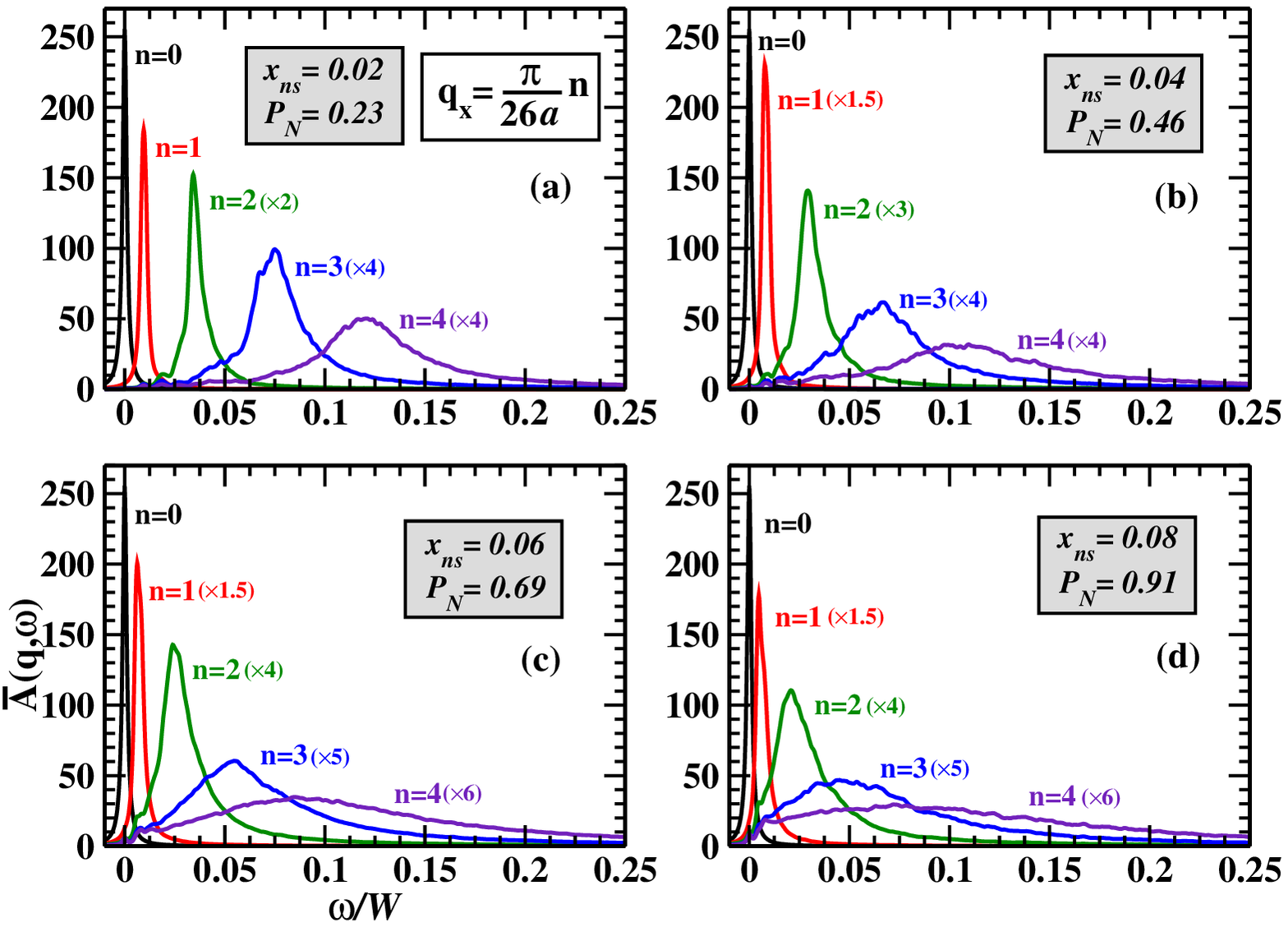}}
\caption{Average spectral function $\bar{A}({\bf q},\omega)$ as a function of the energy in the [1 0 0] direction, for four different $x_{ns}$. 
$P_N$ indicates the percentage of total impurities inside the nanospheres. The parameters are $\lambda$=$a$, $r_0$=$2a$, $x_{in}$=0.8, and $N$=52$^3$. The 
$x$ axis is in units of $\omega/W$. (The intensity of the peaks have been multiplied by the factors indicated in the parentheses). 
}
\label{fig6}
\end{figure}

\begin{figure}[t]\centerline
{\includegraphics[width=4.2in,angle=-90]{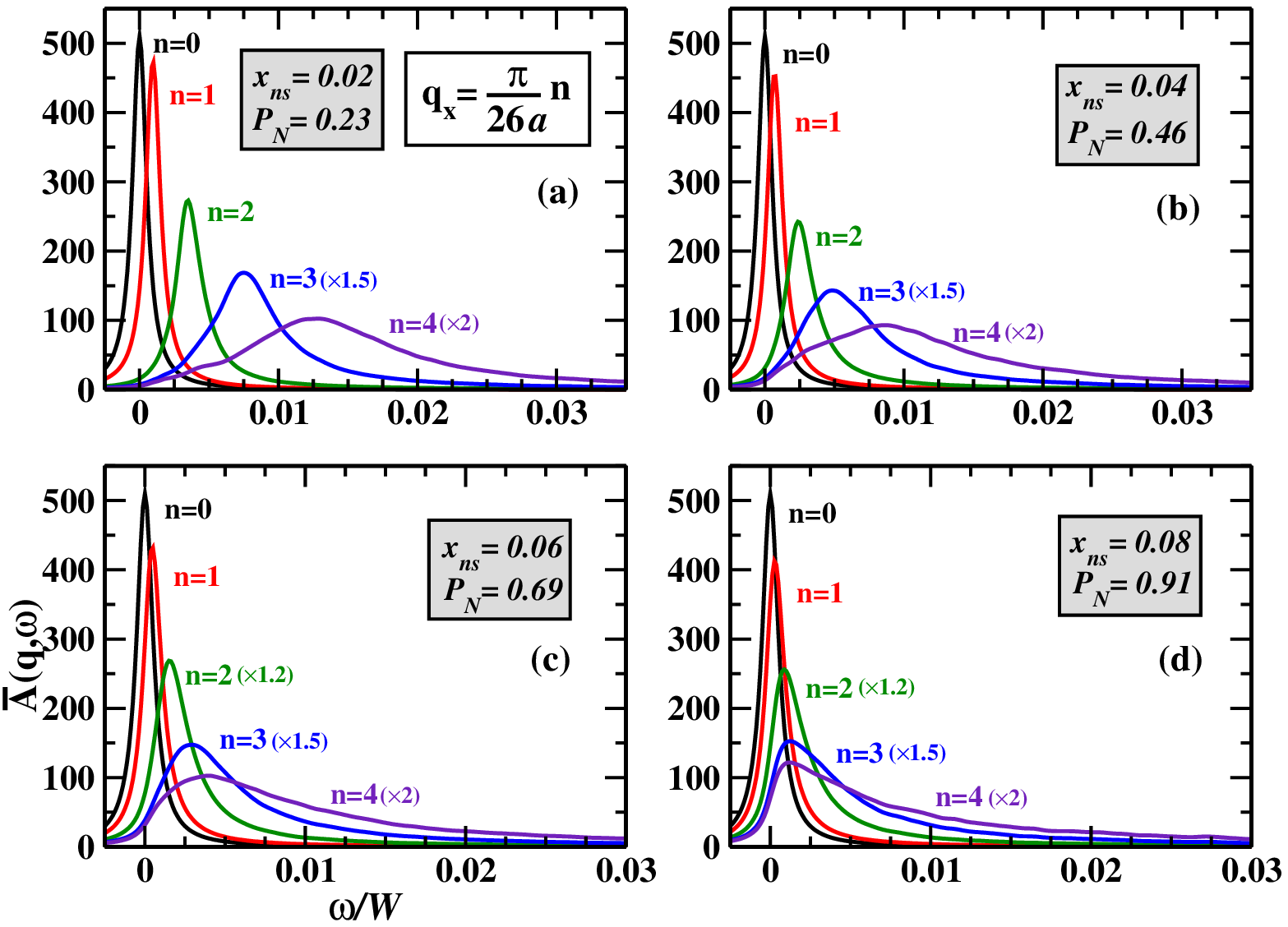}}
\caption{Average spectral function $\bar{A}({\bf q},\omega)$ for the case of $\lambda$=$a/2$, with the other parameters same as in Figure~\ref{fig6}.
}
\label{fig7}
\end{figure}

\ We now proceed further and analyze the effects of nanoscale inhomogeneities on the dynamical spectral function. The results for $\lambda$=$a$ are depicted in Figure~\ref{fig6}. 
As we increase $x_{ns}$, there is a broadening in the excitations, accompanied with an increase in asymmetry  
and a shift toward the lower energies is observed. These effects are already pronounced even 
for the lowest concentration of nanospheres. For instance, for $q_{x}a\approx 0.12\pi$ (n=3), the magnon energies are 0.09$W$, 0.075$W$, and 0.067$W$, for $x_{ns}$=0, 0.02, and 0.04, 
respectively. In order to analyze the effects of inhomogeneities on the magnon lifetime for a given $\bf q$, we define the ratio $R({\bf q})=\gamma({\bf q})/\omega({\bf q})$, 
where $\gamma({\bf q})$ is the half-width of the excitations. The excitations are well-defined in character only when $R({\bf q})<1$. For the aforesaid $q_x$, the corresponding 
$R({\bf q})$'s  are 0.2, 0.33, and 0.66, for $x_{ns}$=0, 0.02, and 0.04, which corresponds to an increase of about 60\% and 200\% respectively, compared to the homogeneous case, 
for these two values of $x_{ns}$. It is interesting to note that these effects could hardly be anticipated from the magnon DOS results (Figure~\ref{fig3}). In fact the 
analyses of the DOS suggested that the low energy excitations should be weakly affected by the inhomogeneities.  
In the following, we discuss the spectral function in the presence of short-ranged interactions, shown in Figure~\ref{fig7}. As in the previous case, well-defined excitations 
exist only for small values of the momentum. However, here we find that the shift toward the lower energies is strongly enhanced. If we consider 
the particular case of $q_{x}a \approx 0.12\pi$, the magnon energies are shifted by 30\% and 60\% respectively, for $x_{ns}$=0.02 and 0.04, with respect to that of 
the homogeneous case. The $R({\bf q})$'s for this value of $q_x$ are 0.4, 0.8, and 1.3 for $x_{ns}$=0, 0.02, 
and 0.04. This indicates that the excitations have dramatically lost their well defined character as compared to the previous case (Figure~\ref{fig6}).

\begin{figure}[htbp]
\centering
\subfigure{
\includegraphics[width=2.55in,angle=0]{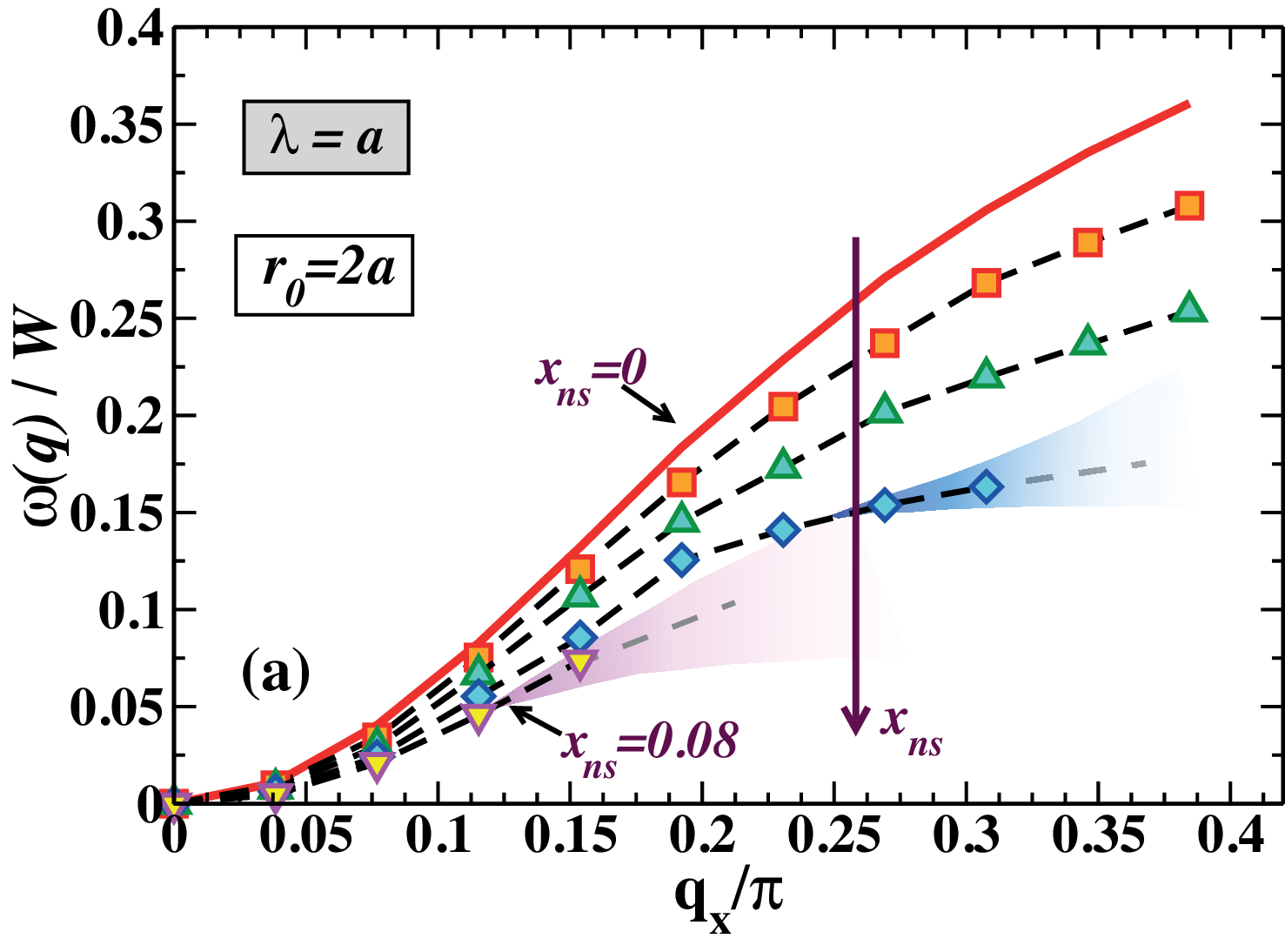}
\label{fig8a}
}
\subfigure{
\includegraphics[width=2.55in,angle=0]{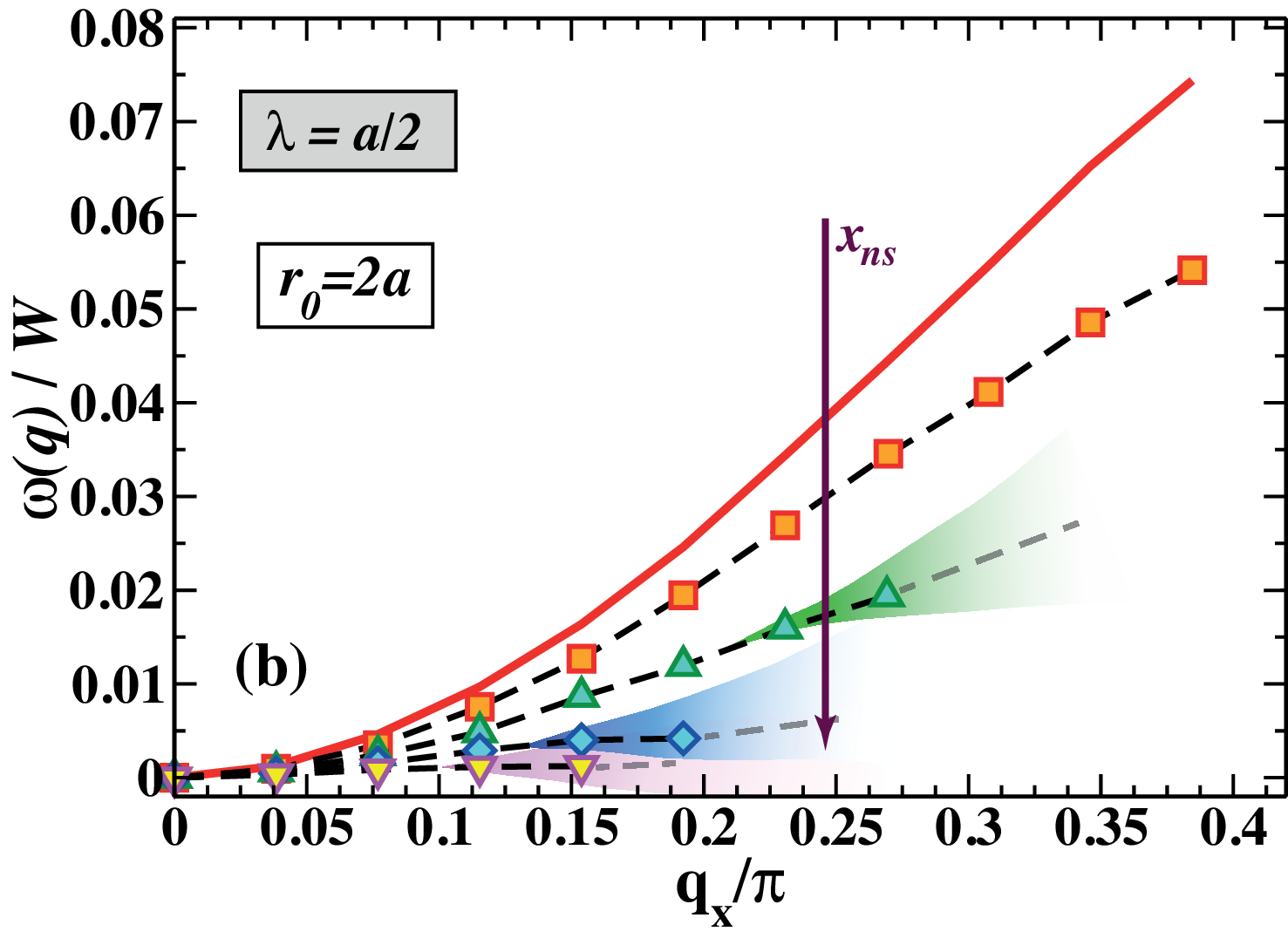}
\label{fig8b}
}
\subfigure{
\includegraphics[width=2.55in,angle=0]{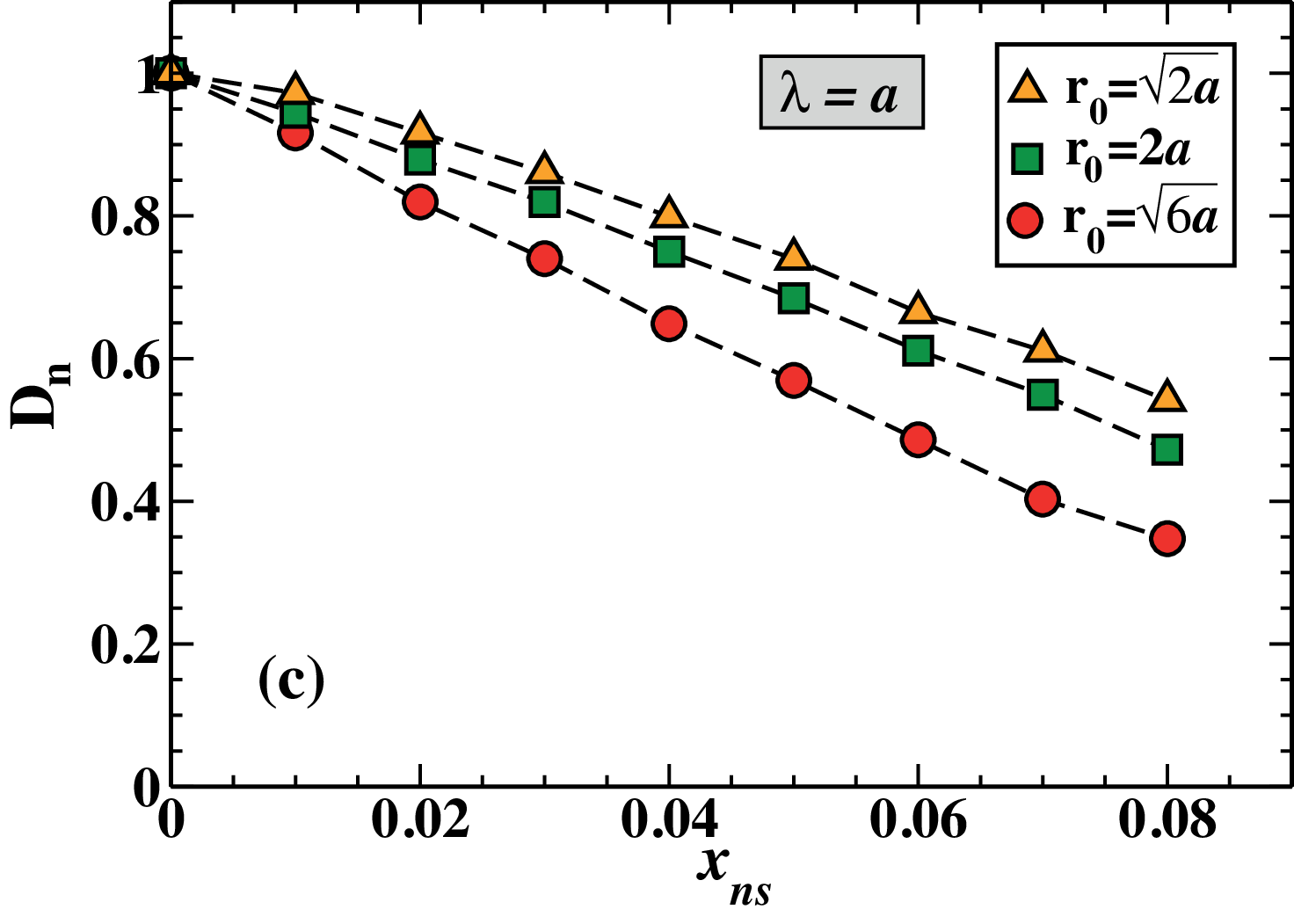}
\label{fig8c}
}
\subfigure{
\includegraphics[width=2.55in,angle=0]{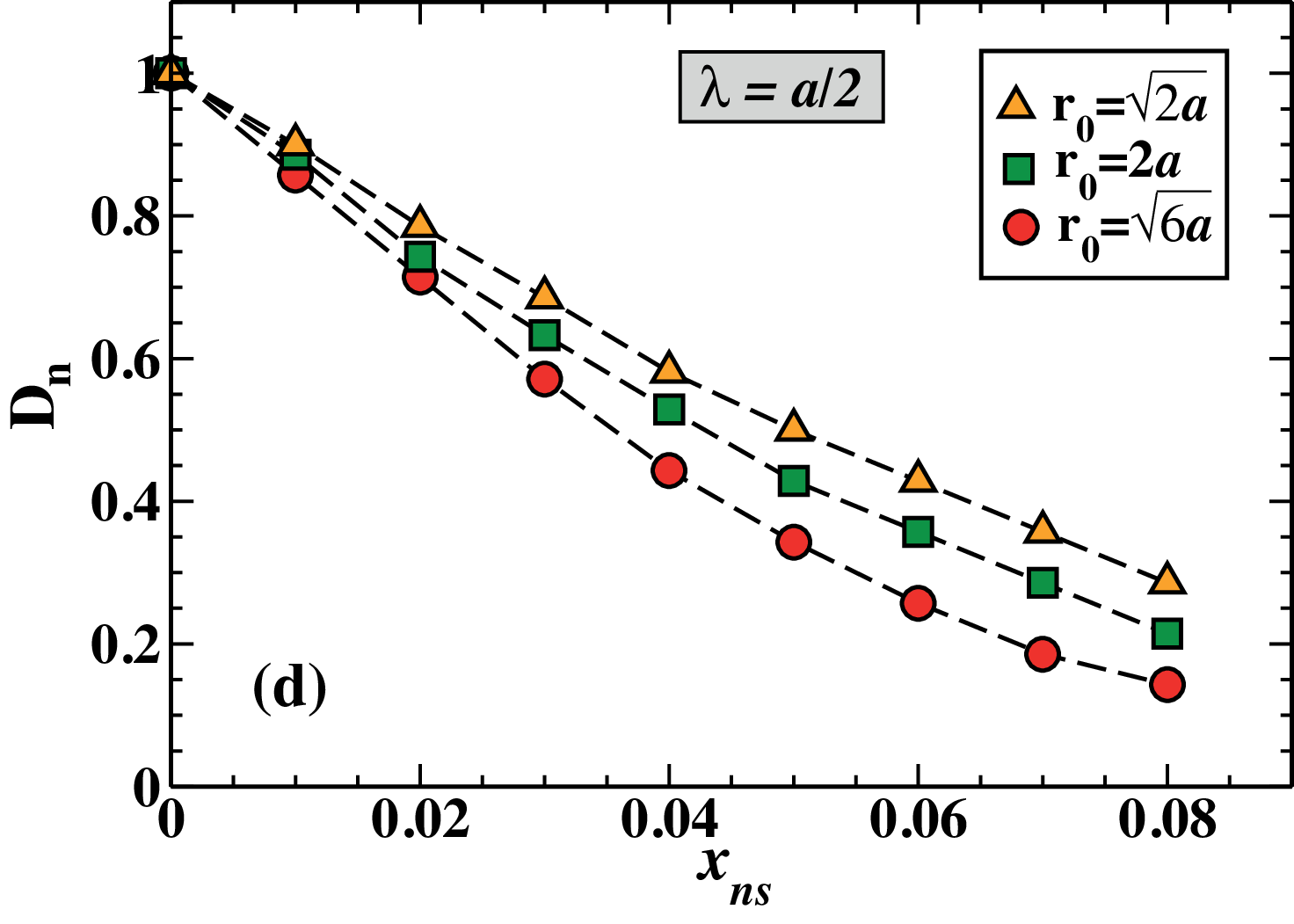}
\label{fig8d}
}
\label{fig8}
\caption[Optional caption for list of figures]{
Magnon dispersion as a function of $q_x/\pi$, for different $x_{ns}$ (0, 0.02, 0.04, 0.06, and 0.08) for (a) $\lambda$=$a$, and (b) $\lambda$=$a/2$.
The  vertical arrows indicate the direction of increasing $x_{ns}$. The solid (red) curve corresponds to the homogeneous case. The shaded regions indicate the $q$ values 
beyond which there is no well-defined excitation. (c) and (d) show the 
normalized spin-stiffness $D_n$ as a function of $x_{ns}$, corresponding to  $\lambda$=$a$, and  $\lambda$=$a/2$, respectively, for 
three different $r_0$: $\sqrt{6}a$, $2a$,  and $\sqrt{2}a$. The concentration inside nanospheres is $x_{in}$=0.8.
}
\end{figure}

\ From the $\bar{A}({\bf q},\omega)$ calculations, we extract both the magnon dispersion $\omega({\bf q})$ and the spin-stiffness ${D^{inh}}$. We remind that in
the long wavelength limit,  $\omega({\bf q})$=$D^{inh}q^2$. In order to gauge the effects of the 
inhomogeneities, we define the normalized spin-stiffness coefficient $D_n$=${D^{inh}}/{D^{hom}}$, where $D^{inh}$ denotes the spin-stiffness of the inhomogeneous 
systems, and $D^{hom}$ that of the homogeneously diluted system. The values of $D^{hom}$ are found to be 2.9$J_0$ and 0.07$J_0$, for $\lambda$=$a$ 
and $a/2$ respectively, for $x$=0.07. We would like to stress that both $\omega(\bf q)$ and ${D^{inh}}$ were obtained from the first non-zero $\bf q$ excitations 
($\bf q$=($\frac{2\pi}{La}$, 0, 0)) corresponding to system sizes $N=L^3$,  where $L$ varies from 36 to 52. 
Figures~\ref{fig8a} and \ref{fig8b} show $\omega(\bf q)$ as a function of $\bf q$ for various $x_{ns}$, corresponding to $\lambda$=$a$, and $a/2$ respectively. 
First, in both cases, we observe that the inhomogeneities suppress the magnon dispersion. This suppression is even more drastic in the case of short-ranged 
interactions. In addition, the quadratic behavior persists up to large $\bf q$ for $\lambda$=$a/2$, whilst a clear softening of the magnon modes can be seen in the 
case of the extended couplings, for $q_x/\pi\ge$0.2. This is consistent with the discussions of Figures~\ref{fig6} and ~\ref{fig7}. 
The well-defined magnon region shrinks significantly on going from $\lambda$=$a$ to $\lambda$=$a/2$. This tendency is enhanced on increasing $x_{ns}$.
With increase in $x_{ns}$, the concentration of impurities outside the clusters gradually decreases since the overall concentration is fixed. Consequently 
the typical distance between the impurities outside increases and they are weakly coupled to the rest of the system. This leads to the suppression of the magnon 
modes with increasing $x_{ns}$ and the effects are naturally more pronounced in the case of short-ranged interactions.  
In order to evaluate more quantitatively the effects of the inhomogeneities, the normalized spin-stiffness $D_n$ is plotted  in Figures~\ref{fig8c} and \ref{fig8d},  
as a function of $x_{ns}$, for various radii of the nanospheres. As anticipated from the previous figures, one observes 
for both $\lambda$=$a$ and $a/2$, a monotonous decrease of the spin-stiffness with increasing $x_{ns}$. $D_n$ decreases almost linearly up to $x_{ns}\approx$0.04, and 
the slope for $\lambda$=$a/2$ is twice that of $\lambda$=$a$. We find that the suppression of the spin-stiffness is greater for larger radii of the nanospheres.
For instance, for the particular $x_{ns}$=0.03, the spin-stiffness is reduced by 15\% for $r_0$=$\sqrt{2}a$, and by almost 
30\% for $r_0$=$\sqrt{6}a$, for $\lambda$=$a$ (Figure~\ref{fig8c}), although the concentration outside the nanospheres is almost the same. In the case of 
the short-ranged couplings this effect is even stronger. The effects of the nanoscale inhomogeneities on the spin-stiffness are in striking contrast to that 
observed on the critical temperatures, where an increase by more than one order of magnitude was found\cite{akash2012_1}.

\section{Conclusion}

\ To conclude, in this work we have shed light on the effects of nanoscale inhomogeneities on the magnetic excitation spectrum of dilute systems. To our knowledge, 
no such detailed numerical study has been performed so far. Compared to homogeneously diluted systems, even low concentrations of nanoscale inhomogeneities is sufficient to 
induce drastic changes in the magnon DOS. The magnon dispersion is found to vary significantly with the concentration of inhomogeneities, leading to a strong suppression of 
the spin-stiffness. 
This suppression of the dispersion can be ascribed to the reduction of the effective interactions with increase in the density of nanoclusters in the system. 
In the rapidly growing field of spintronics, most data storage devices use the dynamical motion of spins. In this context, this study on the spin-wave damping can be 
of practical interest.  
We also find a strong increase of the half-width of the magnon excitations in the long wavelength limit. These features are strongly enhanced in the case of short-ranged couplings. 
The lifetime of the excitations (which is inversely proportional to the half-width) is also relevant from the applications aspect. A short lifetime is essential 
for memory devices, to leave a bit in a steady state after a read-in or read-out operation. A longer lifetime, on the other hand, is necessary for the unhampered transmission of 
signals in inter-chip communications.  
It would also be interesting to study the effects of temperature on the spin dynamics of these inhomogeneous systems. We expect the temperature 
dependence of the spin-stiffness, especially in systems with short-ranged interactions, to be unconventional compared to that of the homogeneous systems. 
Another follow up to this work, would be to calculate the couplings, within a non-perturbative approach, in the presence of the inhomogeneities. The present 
findings already provide qualitative insights and these could be useful in understanding further calculations. Finally, we believe this detailed study would serve to 
motivate future experimental works on these inhomogeneous compounds.

\ack
This work was partially supported by the EU within FP7-PEOPLE-ITN-2008, Grant number 234970 Nanoelectronics: Concepts, Theory and Modeling. AC and PW acknowledge financial 
support by DFG within the collaborative research project SFB 689 C3.

\end{document}